\documentclass[aps,prc,12pt,reprint,amsmath,amssymb,superscriptaddress]{revtex4-2}

\usepackage{graphicx}
\usepackage{multirow}
\usepackage{booktabs}
\usepackage{mathptmx}
\usepackage{microtype}
\usepackage[colorlinks=true,linkcolor=blue,citecolor=blue,urlcolor=blue]{hyperref}

\begin{document}

	\title{Non-locality function of microscopic optical potentials from Skyrme-based nuclear structure models at low energies}

	\author{Do Quang Tam}
	\affiliation{Faculty of Basic Sciences, University of Medicine and Pharmacy, Hue University, Hue City 52000, Vietnam}
	\affiliation{Department of Physics, University of Education, Hue University, Hue City 52000, Vietnam}
	\affiliation{Center for Theoretical and Computational Physics, University of Education, Hue University, Hue City 52000, Vietnam}
	\author{N. Hoang Tung}
	\affiliation{Interdisciplinary Research Center for Industrial Nuclear Energy (IRC-INE), KFUPM, Dhahran, Saudi Arabia}
	\author{N. Hoang Phuc}
	\affiliation{	Department of Applied Physics, Faculty of Applied Science, Ho Chi
		Minh City University of Technology, 268 Ly Thuong Kiet
		Street, Dien Hong Ward, Ho Chi Minh City, Vietnam}
	\affiliation{	Vietnam National University Ho Chi Minh City, Linh Trung Ward, Ho
		Chi Minh City, Vietnam}
	\author{T. V. Nhan Hao}
	\email[Corresponding author: ]{tvnhao@hueuni.edu.vn}
	\affiliation{Department of Physics, University of Education, Hue University, Hue City 52000, Vietnam}
	\affiliation{Center for Theoretical and Computational Physics, University of Education, Hue University, Hue City 52000, Vietnam}

	\date{\today}

\begin{abstract}

The non-locality of the imaginary part of the
microscopic optical potential is investigated for neutron elastic
scattering on $^{16}$O, $^{40}$Ca, $^{48}$Ca, and $^{208}$Pb at
incident energies up to 30~MeV. The potential is derived within the
particle-vibration coupling framework using fully self-consistent
Skyrme-Hartree-Fock plus RPA calculations with the SLy5
interaction, without \textit{ad hoc} adjustable parameters. By fitting the non-locality coordinate dependence of the
absorptive potential $W(R,s) \equiv \mathrm{Im}\,\Delta\Sigma$,
where $\Delta\Sigma$ is the dynamical part of the nucleon
self-energy, to a Gaussian form at representative volume and
surface radial positions, we extract the radius- and energy-dependent non-locality function $\beta(R,E)$, whose values range from $1.01$ to $2.03$~fm, with a clear
dependence on radial position, incident energy, and target mass.  These results demonstrate that, for the absorptive part of the optical potential, a single universal constant such as the
Perey--Buck value $\beta = 0.85$~fm cannot capture the
radial, energy, and target-mass dependence of the non-locality, and that nucleus-dependent,
energy-dependent, and radially-resolved descriptions of the absorptive potential $W$ are
required for precision nuclear reaction calculations.

\end{abstract}

	\maketitle

\section{Introduction}

In nuclear reaction theory, the non-locality parameter \( \beta \) plays a central role in the accurate modeling of nucleon--nucleus interactions within the optical potential framework \cite{deltuva2009three,timofeyuk2013nonlocalityab,timofeyuk2013nonlocality,titus2014testing,fraser2008two,bethe1956nuclear,frahn1957velocity,perey1962non,jaghoub2018exploration,mahzoon2014forging,deltuva2009three,timofeyuk2013nonlocality,titus2016explicit,li2018nonlocal,galetti1998nonlocal,teruya2016nonlocality,perez2019assessment,galetti1994nonlocal,bai2021generalizing}. Its origin can be traced to several microscopic mechanisms, including antisymmetrization of wave functions due to the Pauli exclusion principle, finite-range characteristics of the \( NN \) interaction, channel couplings, and multiple scatterings. A widely used phenomenological representation of non-locality is provided by the energy-independent Perey and Buck potential, which employs a Gaussian non-locality kernel $
e^{-s^2/\beta^2}$
characterized by \textit{ad hoc} non-locality parameter \( \beta = 0.85 \, \mathrm{fm} \), and $s=r-r'$ \cite{perey1962non}. This value was determined by fitting elastic neutron scattering data on \( ^{208}\mathrm{Pb} \) at incident energies of 7 MeV and 14.5 MeV. Despite being constrained by data at only two energies, the Perey--Buck potential offers a remarkably universal framework for describing nucleon--nucleus scattering across a wide range of targets and energies. Once the parameter $\beta=0.85$ fm was fitted from the  \( ^{208}\mathrm{Pb} \) data, it was applied without further adjustment to calculations on a range of nuclei from Al to Pb and at energies from 0.4 MeV to 24 MeV.
This parameter is a key component of the Perey correction factor, which is employed to simplify calculations, particularly within the distorted wave Born approximation (DWBA), by approximating the non-local Schrödinger equation with an equivalent local form. However, the work of L. J. Titus \textit{et al.}~\cite{titus2014testing} has shown that the effects of non-locality cannot be fully captured by the traditional Perey correction factor, highlighting the need for more accurate and systematically derived methods to account for the non-local nature of the nucleon--nucleus interaction. This issue becomes particularly critical in the context of modern nuclear reaction models, such as \((d,p)\) and \((p,d)\) reactions in inverse kinematics, which are essential tools for investigating the structure of exotic nuclei. In the unstable nuclei region, the microscopic optical potential is expected to be a useful tool to provide some guidance about non-locality properties of nucleon--nucleus interaction. Recently, the non-local structure of microscopic optical model potentials  is calculated in momentum space, which gives a non-locality form factor featuring a bell shape with non-locality range between 0.86 and 0.89 fm for both  proton and neutron at incident energies below 65 MeV \cite{arellano2022separability,arellano2024universal}.

Non-locality is not merely a phenomenological artifact but a fundamental and unavoidable feature of microscopic optical potentials derived from many-body nuclear theories.
 Microscopic optical potentials can be derived from a variety of many-body theoretical frameworks that incorporate the complex, non-local, and energy-dependent nature of nucleon–nucleus interactions. Among these, Green’s function theory plays a central role by linking the optical potential to the irreducible nucleon self-energy, naturally embedding non-locality and dispersion effects.
 The dispersive optical model (DOM), which builds upon this connection, imposes dispersion relations to constrain the energy dependence of both real and imaginary components of the potential using experimental data \cite{mahaux1986calculation,dickhoff2021linking,charity2007dispersive,dickhoff2010nonlocal,atkinson2020dispersive}.

 Brueckner–Hartree–Fock (BHF) theory constructs optical potentials from in-medium nucleon–nucleon G-matrix interactions folded over nuclear densities, providing insight into medium modifications and density dependencies \cite{arellano2022separability, rafi2012brueckner}.
 Additionally, \textit{ab initio} approaches including the self-consistent Green's function (SCGF) and coupled-cluster methods yield optical potentials consistent with the same underlying nuclear forces used in structure calculations \cite{rotureau2017optical, burrows2024ab, idini2019ab}.
These frameworks collectively demonstrate that non-locality is a robust and intrinsic feature of the nucleon self-energy and must be incorporated for a consistent description of nuclear reactions.

At low incident energies, where specific nuclear structure effects, such as collective vibrational excitations, play a significant role, N.~Vinh~Mau derived the optical potential \( U(\bold{r}, \bold{r}'; E) \) from the mass operator or self-energy \( \Sigma(\bold{r}, \bold{r}'; E) \) including RPA states, leading to an inherently non-local, energy-dependent, and complex potential that reflects the underlying nucleon correlations and exchange mechanisms within the nucleus~\cite{vinh1970microscopic}. In this framework, a simple zero-range effective nucleon-nucleon interaction was employed to model the imaginary part of the potential for neutron scattering on \( ^{40}\mathrm{Ca} \) at low energies \cite{mau1976optical}. The non-local nature of the resulting microscopic optical potential (MOP) was investigated in both the surface and volume regions of the nucleus.

\begin{figure*}[!ht]
	\includegraphics[width=0.85\linewidth]{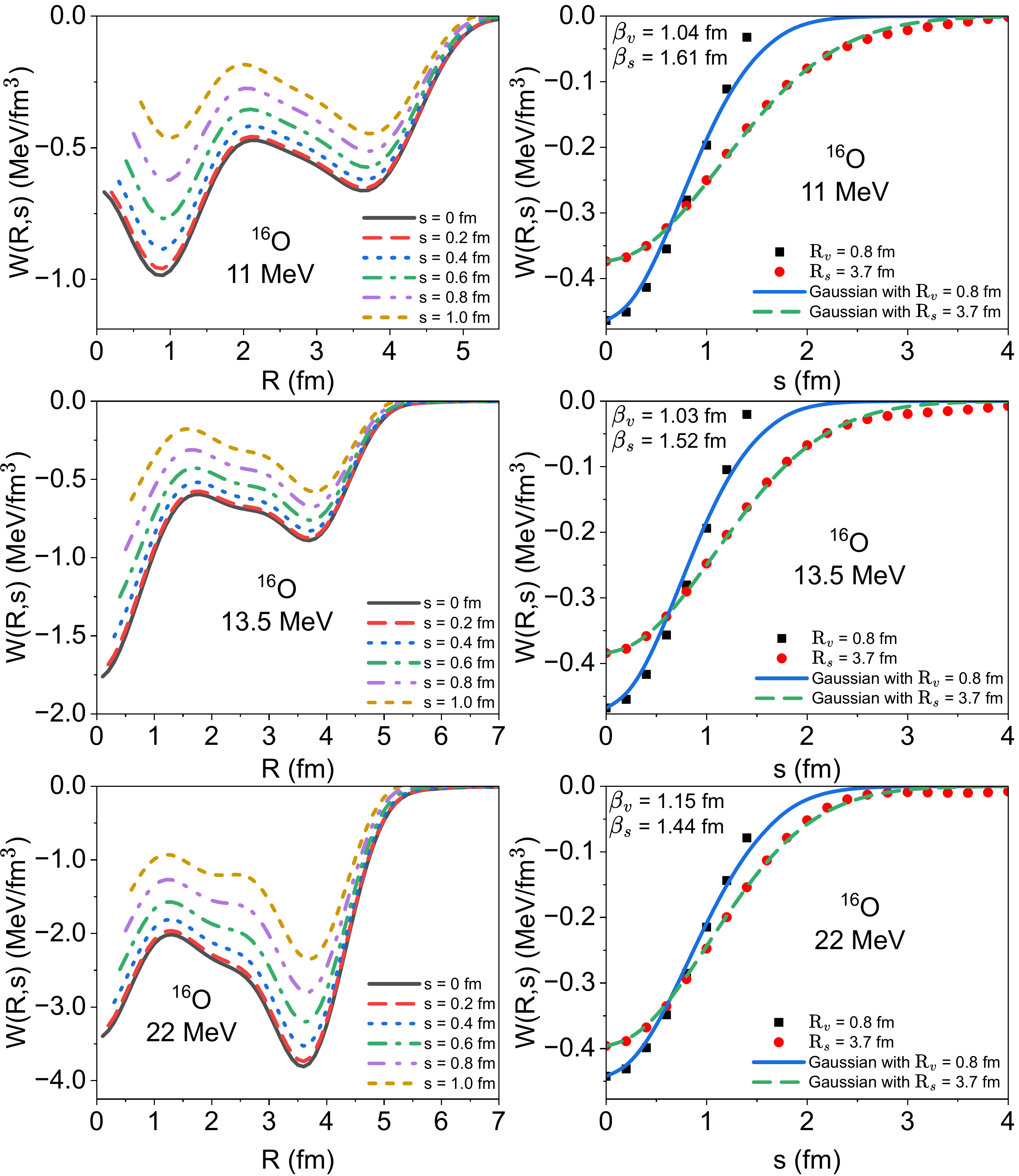}%
	\caption{Imaginary part of the non-local optical potential $W(R, s)$ and extracted non-locality parameter $\beta$ for neutron elastic scattering on $^{16}$O at incident energies of 11~MeV, 13.5~MeV, and 22~MeV. \textbf{Left panels:} $W(R, s)$ as a function of the center-of-mass coordinate $R$ for various fixed non-locality coordinates $s$. \textbf{Right panels:} $W(R,s)$ as a function of the non-locality coordinate $s$ at representative radial positions $R_v$ (black squares, volume region) and $R_s$ (red circles, surface region). The blue solid curve represents the Gaussian fit for the volume region with the extracted $\beta_v$, while the green dashed curve represents the Gaussian fit for the surface region with the extracted $\beta_s$.}
	\label{h1}
\end{figure*}

Later, this approach was extended to the finite-range Gogny effective interactions, which ensure a realistic representation of the nucleon-nucleon interaction, capturing the essential features that contribute to non-locality \cite{blanchon2015microscopic,blanchon2015prospective,blanchon2017asymmetry}. By inherently incorporating non-locality and energy dependence, this approach offers an accurate and physically grounded description of nucleon-nucleus interactions. A Perey-Buck equivalent potential has been fitted by using the microscopic optical potential generated from effective Gogny interaction for \( ^{40}\mathrm{Ca} \) and \( ^{48}\mathrm{Ca} \). These calculations can give valuable guidance for further non-local potential parametrizations.  However, due to the heavy numerical calculation using Gogny interaction, the modest calculations are limited to the region of medium nuclei \cite{blanchon2017asymmetry}.
\begin{figure*}[!ht]
	\includegraphics[width=0.85\linewidth]{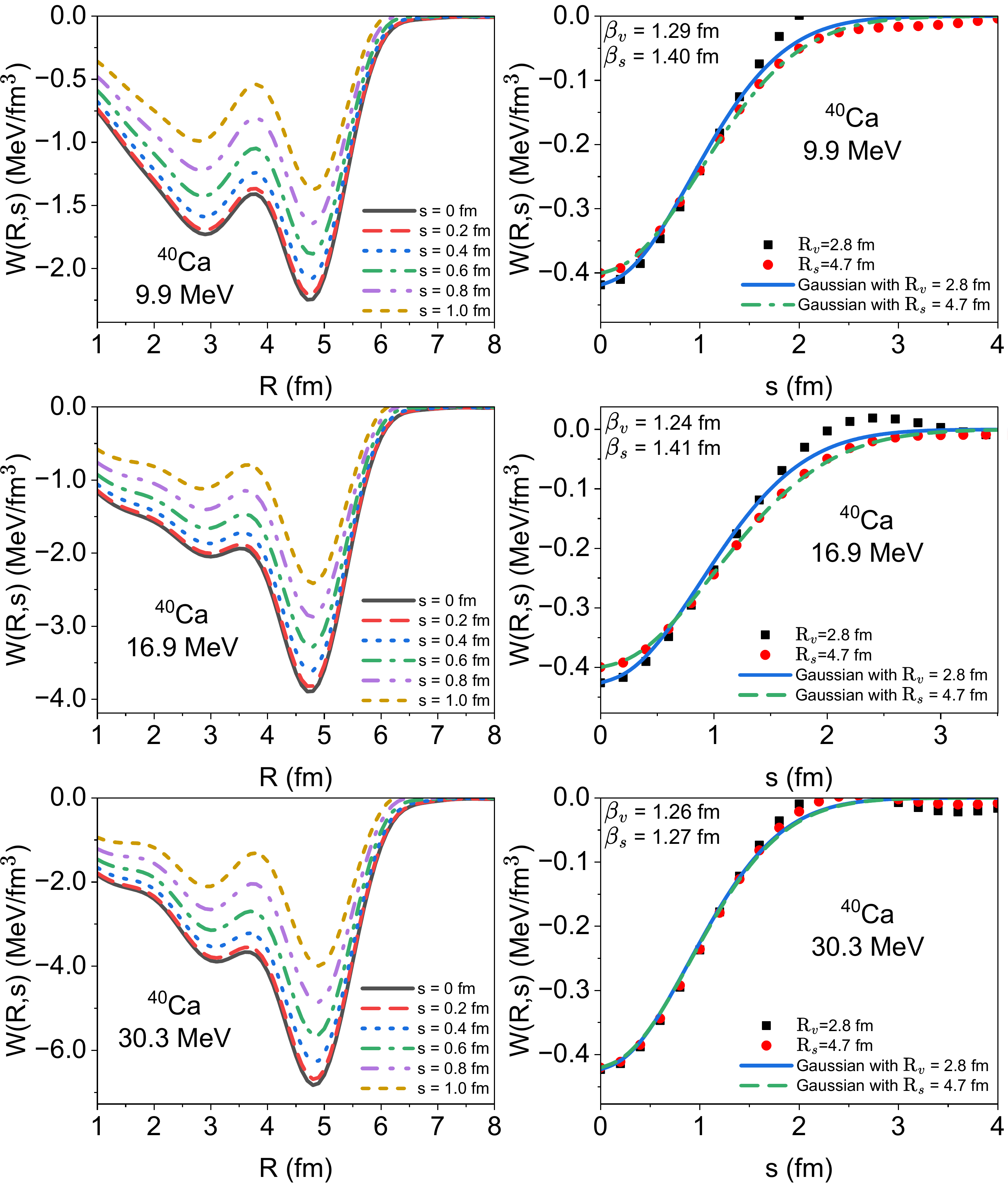}%
	\caption{Same as FIG.~1, but for $^{40}$Ca at incident energies of 9.9~MeV, 16.9~MeV, and 30.3~MeV.}
	\label{h2}
\end{figure*}

Concurrently with the calculations using the Gogny interaction, similar studies were conducted by seminal work by V. Bernard \textit{et al.}\cite{bernard1979microscopic}, employing Skyrme effective zero-range interactions. The non-locality parameter $\beta$ is also extracted by the Gaussian shape of the absorption part of the microscopic optical potential: $W(R,s)=W(R,s=0) \times e^{-s^2/\beta^2} $, where $R=\frac{r+r'}{2}$. By performing the calculations for neutron elastic scattering off $^{208}\mathrm{Pb}$ at energies below 30~MeV, they found $\beta_s$ varies slightly from 0.98 fm at $E=5$ MeV to 0.92 fm at $E = 30$ MeV only at the peak radius $R=7.2$~fm on the surface. In the nuclear interior, it is not possible to extract the non-locality parameter $\beta_v$ due to the absence of absorptive strength.

Based on the work of Ref. \cite{bernard1979microscopic}, recent advances in computational power and many-body theoretical methods \cite{colo2010effect,colo2013self,colo2020nuclear} allow us to use the effective Skyrme interaction consistently at all levels, from the mean-field description to the treatment of correlations beyond the mean field in a model space which includes all natural parity RPA excited states with multipolarity $L=0^{+},1^{-},2^{+},3^{-},4^{+},5^{-}$ and excitation energies below 50 MeV.  In conjunction with approaches employing the Gogny interaction, these developments offer a unified and self-consistent framework for describing nuclear structure and reactions at low energies. In comparison with the results of Ref. \cite{bernard1979microscopic}, the calculated absorption is not only enhanced at the nuclear surface but also exhibits an important contribution in the nuclear interior for all double closed shell nuclei $^{16}$O,$^{40}$Ca, $^{48}$Ca, and $^{208}$Pb at incident energies below 50~MeV\cite{hao2015low,nhan2018microscopic,hoang2020effects}. Building on this progress, as a first step, we focus on investigating the non-locality parameter \( \beta \) from the nuclear interior to the surface for a range of target nuclei, including \( ^{16}\mathrm{O} \), \( ^{40}\mathrm{Ca} \), \( ^{48}\mathrm{Ca} \), and \( ^{208}\mathrm{Pb} \), at incident energies below 50~MeV.
\begin{figure*}[!ht]
	\includegraphics[width=0.85\linewidth]{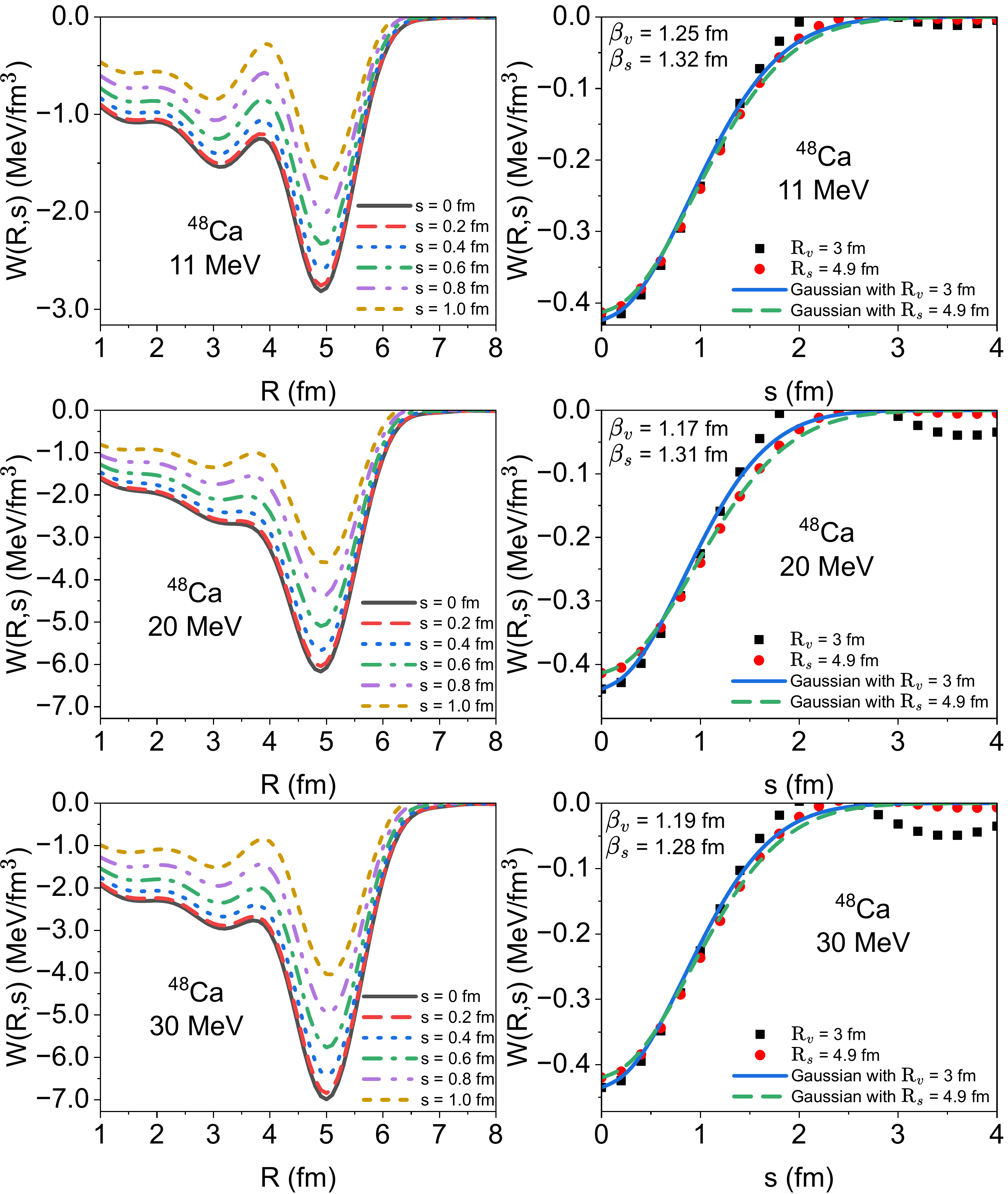}%
	\caption{Same as FIG.~1, but for $^{48}$Ca at incident energies of 11~MeV, 20~MeV, and 30~MeV.
	}
	\label{h3}
\end{figure*}

\section{Theoretical framework}

According to Refs. \cite{mau1976optical,bernard1979microscopic,hao2015low,nhan2018microscopic,hoang2020effects},  we define the quantity $W(\mathrm{\bold{R}},\mathrm{\bold{s}},E) \equiv \textrm{Im}\Delta\Sigma(\mathrm{\bold{r}},\mathrm{\bold{r'}},E)$ as a function of $ \mathrm{\bold{R}} = \frac{1}{2} (\mathrm{\bold{r}} + \mathrm{\bold{r'}}) $, which corresponds to the radius and shape of $\textrm{Im}\Delta\Sigma$, and $\mathrm{\bold{s}} = \mathrm{\bold{r}} - \mathrm{\bold{r'}}$, which characterizes its non-locality.  The $\Delta\Sigma(\mathrm{\bold{r}},\mathrm{\bold{r'}},E)$ is non-local, complex, and energy dependent giving a major contribution to the imaginary part which is responsible for the absorption of the optical potential:
\begin{equation}
	V_{\mathrm{opt}}(\mathrm{\bold{r}},\mathrm{\bold{r'}},E)=V_{\mathrm{HF}}(\mathrm{\bold{r}})+\Delta \Sigma(\mathrm{\bold{r}},\mathrm{\bold{r'}},E),
	\label{2}
\end{equation}
where
\begin{equation}
	\Delta \Sigma(\mathrm{\bold{r}},\mathrm{\bold{r'}},E)=\Sigma(\mathrm{\bold{r}},\mathrm{\bold{r'}},E)-\frac{1}{2} \Sigma^{(2)}(\mathrm{\bold{r}},\mathrm{\bold{r'}},E).
	\label{3}
\end{equation}
In the general case $W$ is a very complicated function of $r$, $r'$ and $\theta$ the angle between the two vectors  $\mathrm{\bold{r}}$, and $\mathrm{\bold{r'}}$.

The potential can be expanded as:
\begin{equation}
	W(\mathrm{\bold{R}},\mathrm{\bold{s}},E) = \sum_{\Lambda=0}^{\infty} W_{\Lambda}(s, R,E) P_{\Lambda}(\cos\theta),
\end{equation}
where $P_{\Lambda}$ are Legendre polynomials. Vinh Mau and Bouyssy \cite{mau1976optical} demonstrated that the term $\Lambda = 2$ of the potential, and \textit{a fortiori} terms with $\Lambda > 2$, is much smaller than the first term $\Lambda = 0$, at least by an order of magnitude. Therefore, to a very good approximation one can write $	W(\mathrm{\bold{R}},\mathrm{\bold{s}},E) = 	W_0(R,s,E)$.
This independence of $W$ upon $\theta$ justifies treating $R = |\bold{R}|$ and $s = |\bold{s}|$ as scalar quantities (magnitudes) in our analysis. This is the standard convention universally adopted in phenomenological non-local optical potentials, including the seminal Perey-Buck potential~\cite{perey1962non} as well as the Engelbrecht-Fiedeldey parametrization~\cite{engelbrecht1967nonlocal}. In both cases, $s$ and $R$ are explicitly treated as scalar variables representing the magnitudes of the relative and center-of-mass coordinates, respectively.

In Eq. \eqref{2}, $V_{\mathrm{HF}}$ is a static Skyrme-Hartree-Fock mean field which is real, local, and energy independent, providing a major contribution to the real part of the optical potential. To take into account the issue of the Pauli principle correction, $\Sigma^{(2)}(E)$ is the second-order potential (SOP) generated from uncorrelated particle-hole contribution, and $E$ is the nucleon incident energy. When a zero-range interaction such as the Skyrme force is employed, the non-locality of the optical potential originates exclusively from the dynamical component of the self-energy, $\Delta\Sigma(E)$. This contrasts with the case of finite-range interactions, such as the Gogny force, where non-local features are present in both the static mean-field contribution, $V_{\mathrm{HF}}$, and the polarization term $\Delta\Sigma(E)$. However, the use of zero-range Skyrme interactions significantly decreases the numerical complexity of the calculations, which nevertheless remain feasible for nuclei ranging from light to heavy targets. The effective SLy5 Skyrme interaction \cite{chabanat1998skyrme} has been chosen.

Following the convention adopted in Refs.~\cite{bernard1979microscopic}, the dependence of the absorptive potential $W(R,s,E)$ on the non-locality coordinate $s$, at a fixed radial position $R$ and incident energy $E$, is represented by the Gaussian form
\begin{equation}
	W(R,s,E) = W(R,s=0,E)\,
	\exp\!\left[-\frac{s^{2}}{\beta^{2}(R,E)}\right],
	\label{eq:gauss}
\end{equation}
where the width $\beta(R,E)$ characterizes the curvature of $W(R,s)$ as a function of $s$ at fixed $R$ and $E$. Hence $\beta$ is not a single universal constant but a radius- and energy-dependent function, $\beta \equiv \beta(R,E)$. Throughout this work we keep the conventional name ``non-locality parameter'', in line with the Perey--Buck terminology~\cite{perey1962non}, to denote the value of this function at a given $(R,E)$; these values are reported at the representative volume ($R_v$) and surface ($R_s$) radii for incident energies between 5 and 30~MeV. We also note that the present analysis is restricted to the imaginary part of the dynamical self-energy, $W \equiv \mathrm{Im}\,\Delta\Sigma$, which governs the absorption of flux from the elastic channel; the non-locality of the real dynamical component $\mathrm{Re}\,\Delta\Sigma$ is not addressed here, and all statements on the non-locality made in this work therefore refer to the absorptive part only.

\begin{figure*}[!ht]
	\includegraphics[width=0.85\linewidth]{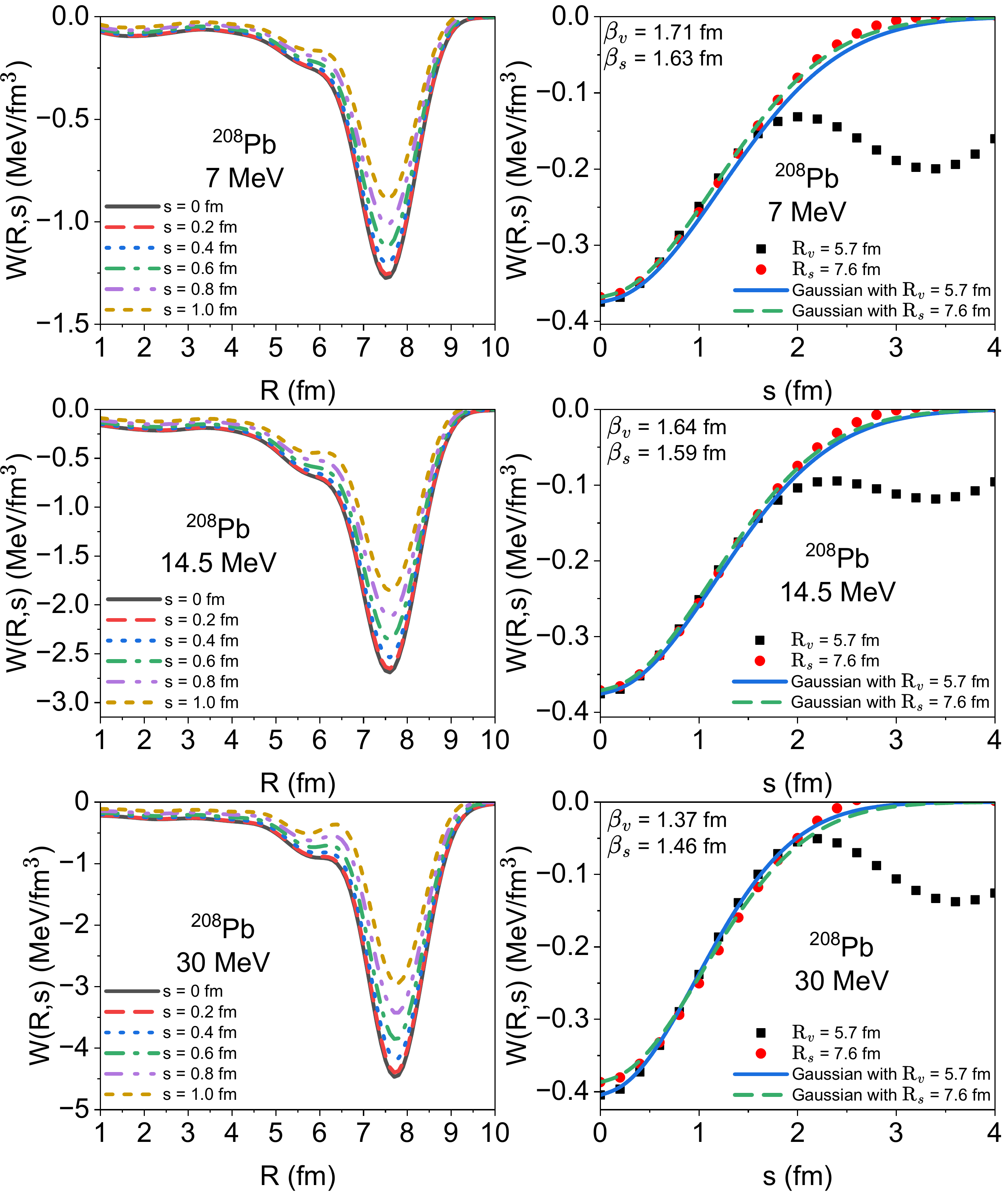}%
	\caption{Same as FIG.~1, but for $^{208}$Pb at incident energies of 7~MeV, 14.5~MeV, and 30~MeV.
	}
	\label{h4}
\end{figure*}

For the description of doubly magic nuclei,  spherical symmetry is adopted to simplify the numerical calculations. First, we solved the radial Hartree-Fock (HF) equations in coordinate space by using a mesh size of 0.1~fm, with the radial coordinate extending up to 15~fm.  Ground and excited states are constructed on top of the HF solution using fully self-consistent RPA calculations~\cite{colo2013self}. All natural-parity RPA excited states with multipolarities \( L = 0^{+}, 1^{-}, 2^{+}, 3^{-}, 4^{+}, 5^{-} \) and excitation energies below 50~MeV are selected for the PVC calculations \cite{colo2010effect,colo2013self}. Continuum states with positive energy are treated by discretizing the spectrum using box boundary conditions. All computational parameters are kept fixed for all nuclei and excitation energies considered, following the methodology of Refs. \cite{nhan2018microscopic,hao2015low,hoang2020effects}. It is worth noting that several numerical codes are now available
for solving the integro-differential Schr\"odinger equation with
non-local optical potentials, as reviewed by Hebborn
\textit{et al.}~\cite{Hebborn2023}. Among these, the code
SIDES~\cite{SIDES2020} provides a dedicated solver for non-local
potentials in coordinate space. In our previous works \cite{hao2015low,nhan2018microscopic,hoang2020effects},  the non-locality
of the microscopic optical potential is treated exactly by employing the distorted wave Born approximation (DWBA) code developed by Raynal~\cite{raynal1998computer}, which solves the full integro-differential equation for the scattering wave function in the presence of a non-local potential.

To keep the present study self-contained and to illustrate the quality of the underlying microscopic potential, Fig.~\ref{fig:obs} (adapted from Ref.~\cite{hao2015low}) shows the neutron elastic-scattering differential cross section for $^{208}$Pb at $E = 14.5$~MeV, obtained from the same fully self-consistent particle-vibration-coupling optical potential whose non-locality is characterized in this work. The scattering observables are computed by solving the non-local Schr\"odinger equation exactly with the DWBA98 code~\cite{raynal1998computer}, without introducing any local-equivalent (Perey-type) approximation. The figure displays the cumulative contributions of the collective multipole modes, obtained by progressively increasing the maximum multipolarity $L_{\max}$ from $1$ to $5$, together with the static Hartree--Fock (HF) result and the experimental data. The dominant contributions to the absorption originate from the most collective $3^{-}$, $4^{+}$, and $2^{+}$ states, whereas the $1^{-}$ and $5^{-}$ modes play a minor role; once all natural-parity modes are included, the measured angular distribution is well reproduced. A systematic account of the angular distributions for all four target nuclei $^{16}$O, $^{40}$Ca, $^{48}$Ca, and $^{208}$Pb over the energy range considered here was reported in our previous works~\cite{hao2015low,nhan2018microscopic,hoang2020effects}.

\section{Results and discussion}

We begin by calculating the imaginary part of the non-local optical potential, \( W(R, s) \), as a function of the radial coordinate \( R \), for fixed non-locality coordinate \( s = 0,0.2, 0.4, 0.6, 0.8, 1.0 \) fm. These calculations are performed for \( ^{16}\mathrm{O} \), \( ^{40}\mathrm{Ca} \), and \( ^{48}\mathrm{Ca} \) and $^{208}$Pb nuclei at various incident energies. As illustrated in Figs. \ref{h1}, \ref{h2}, \ref{h3}, \ref{h4}, the imaginary part \( W(R, s) \), which characterizes the absorptive strength of the potential, systematically decreases with increasing non-locality coordinate \( s \). For the light nucleus \(^{16}\mathrm{O}\) and the medium-mass nuclei \(^{40}\mathrm{Ca}\) and \(^{48}\mathrm{Ca}\), the imaginary part of the optical potential exhibits distinct volume and surface components at all investigated incident energies, with the exception of \(^{16}\mathrm{O}\) at 13.5 and 22~MeV. For heavy nuclei $^{208}$Pb, the surface potential is dominant, except at high energies E=30 MeV for $s=0.8, 1.0$ fm where a minimum appears in the volume region. As a function of $s$, $W(R,s)$ is close to a Gaussian for all incident energies and radii considered.

 We note that the Gaussian approximation holds reasonably well at the nuclear surface for the low-energy regime considered in this work, consistent with the findings of Blanchon \textit{et al.}~\cite{blanchon2015prospective}, who showed good agreement between the non-locality of their microscopic NSM/Gogny potential and a Gaussian shape with $\beta \approx 0.94$~fm at $r = 4.3$~fm for neutron scattering from $^{40}$Ca at 9.91~MeV. However, we acknowledge that a purely Gaussian non-locality does not capture the full complexity revealed by microscopic studies. In particular, Arellano and Blanchon~\cite{arellano2024universal} disclosed a universal separable structure of the optical potential in which the non-locality form factor is inherently complex and of bell-shape (hydrogenic) type, decaying as $e^{-s/b}$ rather than the Gaussian $e^{-s^2/\beta^2}$. This bell-shape non-locality was found to be a robust feature across targets ($^{40}$Ca, $^{90}$Zr, $^{208}$Pb) and beam energies (40--400~MeV), exhibiting a slower falloff and longer tail compared to the Gaussian form. Furthermore, Arellano and Blanchon~\cite{Arellano2018} demonstrated that the non-local structure in coordinate space is not unique: different momentum cutoffs applied to the same momentum-space potential yield different-looking non-localities while preserving identical scattering observables, implying that shape comparisons must be interpreted with caution. The approximate Gaussian behavior found in our Skyrme-based calculations at low energies ($E < 30$ MeV) is therefore understood as a valid surface-dominated description within the energy range studied, while a more complete characterization of the non-locality, particularly in the nuclear interior and at higher energies, would require adopting the bell-shape form factor or similar non-Gaussian prescriptions.

To quantify the quality of the Gaussian fit of $W(R,s)$ from which $\beta$ is extracted, we define the coefficient of determination, denoted $\mathcal{R}^2$, which measures how well the Gaussian model explains the calculated microscopic data. The $\mathcal{R}^2$ value is computed using:

\begin{equation}
	\mathcal{R}^2 = 1 - \frac{\mathrm{SS_{res}}}{\mathrm{SS_{tot}}},
	\label{eq:rsquared}
\end{equation}
where
\begin{equation}
	\mathrm{SS_{res}} = \sum_{i=1}^{N} \left[W_{\mathrm{calc}}(R,s_i)
	- W_{\mathrm{fit}}(R,s_i)\right]^2
	\label{eq:ssres}
\end{equation}
is the residual sum of squares, and
\begin{equation}
	\mathrm{SS_{tot}} = \sum_{i=1}^{N} \left[W_{\mathrm{calc}}(R,s_i)
	- \bar{W}\right]^2,
	\label{eq:sstot}
\end{equation}
is the total sum of squares. Here, $W_{\mathrm{calc}}(R,s_i)$ denotes the calculated $W$ at the fixed radial position $R$ and non-locality coordinate $s_i$,
$W_{\mathrm{fit}}(R,s_i)$ is the
corresponding fitted Gaussian value, $\bar{W} = N^{-1}\sum_{i=1}^{N}
W_{\mathrm{calc}}(R,s_i)$ is the mean of the calculated data at that radius, and $N$ is
the number of data points. Table~\ref{table} summarizes the extracted non-locality parameters $\beta$ along with the corresponding $\mathcal{R}^2$
values for all target nuclei and incident energies presented in
Figs.~\ref{h1}--\ref{h4}. The obtained values of
$\mathcal{R}^2$ range from 0.97 to 0.99, confirming that the
Gaussian ansatz of Eq.~(\ref{eq:gauss}) provides a very good
representation of $W$ across all systems and energies considered.
The fits are systematically better in the surface region
($\mathcal{R}^2 \geq 0.99$) than in the nuclear interior,
where minor deviations from the Gaussian form appear at large
values of the non-locality coordinate $s$. Our calculations with zero-range Skyrme interactions confirm that the Gaussian shape of the imaginary-potential non-locality is independent of the form factors of the underlying interaction.
\begin{table}[htbp]
	\centering
	\caption{Extracted non-locality parameters $\beta$ (in fm) and fit quality $\mathcal{R}^2$ for neutron elastic scattering on $^{16}$O, $^{40}$Ca, $^{48}$Ca, and $^{208}$Pb at various incident energies.}
	\label{table}
	\small
	\begin{tabular}{lcccccc}
		\toprule
		\toprule
		Nucleus & $E$ (MeV)  & $\beta_v$ (fm) & $\mathcal{R}^2_v$  & $\beta_s$ (fm) & $\mathcal{R}^2_s$ \\
		\midrule
		\midrule
		\multirow{3}{*}{$^{16}$O}
		& 11.0  & 1.04 & 0.981  & 1.61 & 0.999 \\
		& 13.5   & 1.03 & 0.976  & 1.52 & 0.998 \\
		& 22.0   & 1.15 & 0.993 & 1.44 & 0.999 \\
		\midrule
		\multirow{3}{*}{$^{40}$Ca}
		& 9.9    & 1.29 & 0.992  & 1.40 & 0.998 \\
		& 16.9  & 1.24 & 0.994 & 1.41 & 0.999 \\
		& 30.3   & 1.26 & 0.992  & 1.27 & 0.997 \\
		\midrule
		\multirow{3}{*}{$^{48}$Ca}
		& 11.0 & 1.25 & 0.995  & 1.32 & 0.998 \\
		& 20.0   & 1.17 & 0.979& 1.31 & 0.998 \\
		& 30.0   & 1.19 & 0.970  & 1.28 & 0.997 \\
		\midrule
		\multirow{3}{*}{$^{208}$Pb}
		& 7.0   & 1.71 & 0.971  & 1.63 & 0.999 \\
		& 14.5  & 1.64 & 0.994   & 1.59 & 0.998 \\
		& 30.0   & 1.37 & 0.999   & 1.46 & 0.996 \\
		\bottomrule
		\bottomrule
	\end{tabular}
\end{table}

\begin{figure*}[!ht]
	\includegraphics[width=0.85\linewidth]{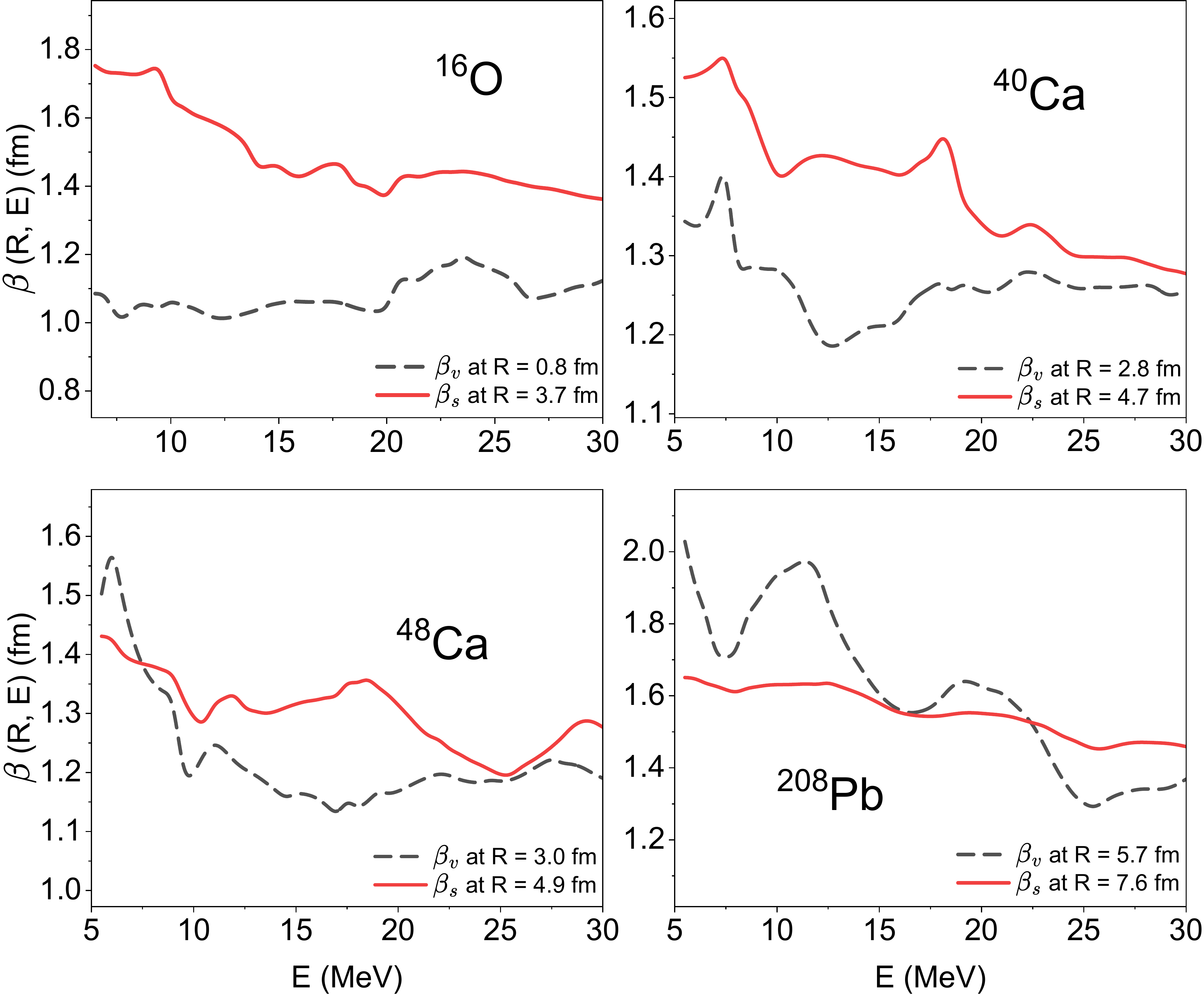}%
	\caption{Energy dependence of the non-locality parameter $\beta$ for neutron elastic scattering on $^{16}$O, $^{40}$Ca, $^{48}$Ca, and $^{208}$Pb over the incident energy range $E = 5$--30~MeV. The gray dashed line represents the volume non-locality parameter $\beta_v$, while the red solid line represents the surface non-locality parameter $\beta_s$.}
	\label{h5}
\end{figure*}

Figure~\ref{h5} presents the energy dependence of the
non-locality parameter $\beta$ for neutron elastic scattering on
$^{16}$O, $^{40}$Ca, $^{48}$Ca, and $^{208}$Pb, evaluated at
representative radial positions corresponding to the nuclear interior
($\beta_{\mathrm{v}}$) and surface ($\beta_{\mathrm{s}}$)
regions, over the incident energy range $E = 5$--$30$~MeV.

For the light nucleus $^{16}$O, a pronounced separation between surface
and volume non-locality is observed across the entire energy range.
The surface parameter $\beta_{\mathrm{s}}$ at $R = 3.7$~fm
decreases from approximately $1.75$~fm at $E \approx 6.5$~MeV to
about $1.36$%
~fm at $E \approx 30$~MeV, exhibiting a monotonic
decreasing trend with a total variation of roughly $22.4\%$. In
contrast, the volume parameter $\beta_{\mathrm{v}}$ at
$R = 0.8$~fm remains comparatively stable around $1.01$%
--$1.20$~fm,
with weak fluctuations and no pronounced energy trend.
The gap $\Delta \beta = \beta_{\mathrm{s}} - \beta_{\mathrm{v}}$
ranges from approximately $0.24$ to $0.72$~fm with a typical
value of about $0.37$~fm, indicating that the non-local
structure in $^{16}$O is governed by qualitatively different
mechanisms in the two regions: the interior is dominated by
mean-field correlations and Pauli blocking, which constrain the
non-locality range, while the surface is more sensitive to
particle-vibration coupling with low-lying collective states,
leading to a substantially broader non-local extent.

For $^{40}$Ca, both $\beta_{\mathrm{v}}$ ($R = 2.8$~fm) and
$\beta_{\mathrm{s}}$ ($R = 4.7$~fm) exhibit an initial peak
near $E \approx 6$--$7$~MeV, reaching $\approx 1.43$%
~fm and
$\approx 1.56$~fm respectively%
, followed by a gradual decrease toward
higher energies. Above $E \approx 20$~MeV, both parameters
stabilize around $1.25$--$1.34$~fm. The difference between surface
and volume values is significantly reduced compared to $^{16}$O,
with $\Delta\beta \lesssim 0.24$~fm over the full energy range
considered, converging to $\Delta\beta \approx 0.01$~fm at
$E = 30.3$~MeV where $\beta_{\mathrm{v}} = 1.26$~fm and
$\beta_{\mathrm{s}} = 1.27$~fm. This near-degeneracy at higher
energies reflects the more uniform nuclear density profile and
stronger shell closure in $^{40}$Ca, which produces a more
homogeneous non-local environment throughout the nuclear volume.

It is instructive to compare these results with the pioneering
microscopic calculation of Vinh~Mau and Bouyssy~\cite{mau1976optical},
who investigated the non-locality of the imaginary optical potential
for neutron-$^{40}$Ca scattering using a zero-range effective
nucleon-nucleon interaction within the Green function formalism
including RPA correlations. In their work, the non-locality range
was evaluated at $R = 1$~fm for the nuclear interior and $R = 3.5$~fm
for the surface, yielding $\beta_{\mathrm{v}} \approx 1$~fm
(nearly energy-independent) and $\beta_{\mathrm{s}}$ decreasing
from $1.58$~fm to $1.36$~fm as $E_0$ increases from 8 to 25~MeV.

In the present calculation, the volume non-locality
$\beta_{\mathrm{v}}$ is evaluated at $R = 2.8$~fm and ranges from
$1.24$ to $1.29$~fm at the energies shown in Table~\ref{table},
reaching $\approx 1.43$%
~fm only at the lowest energies
($E \approx 6$~MeV) as displayed in Fig.~\ref{h5}. The somewhat
larger $\beta_{\mathrm{v}}$ compared to the value of $\approx 1$~fm
reported in Ref.~\cite{mau1976optical} may be attributed to the
different radial positions at which the non-locality is probed,
our evaluation at $R = 2.8$~fm lies closer to the surface region
of $^{40}$Ca than the deep interior point $R = 1$~fm used in
Ref.~\cite{mau1976optical}, where the density is higher and more
uniform and to the fully self-consistent treatment and the
substantially larger RPA model space employed here. Nevertheless, the nearly
energy-independent character of $\beta_{\mathrm{v}}$ reported
in Ref.~\cite{mau1976optical} is qualitatively confirmed by our
results, which show only a modest $\approx 3.3\%$ variation (from $1.29$
to $1.24$~fm) over the energy range $E = 9.9$--$30.3$~MeV.

The most notable difference concerns the surface non-locality
$\beta_{\mathrm{s}}$. Our values, ranging from $1.27$ to
$1.41$~fm, are of the same order as, although still somewhat smaller
than, the $1.36$--$1.58$~fm
reported in Ref.~\cite{mau1976optical}. Moreover, the energy
dependence of $\beta_{\mathrm{s}}$ in our calculations is
considerably weaker: the total variation is about  $9.1\%$ (from
$1.40$ to $1.27$~fm over $E = 9.9$--$30.3$~MeV), compared to
the $\approx 14\%$ decrease (from $1.58$ to $1.36$~fm over
$E_0 = 8$--$25$~MeV) found in Ref.~\cite{mau1976optical}.
Several factors may account for the remaining difference. First, the
surface evaluation radii differ: our $R_{\mathrm{s}} = 4.7$~fm
lies further from the nuclear center than the $R = 3.5$~fm used
in Ref.~\cite{mau1976optical}, which was constrained by the
smaller harmonic oscillator parameter ($b_0 = 1.89$~fm,
corresponding to a nuclear radius of only $3.2$~fm) employed
in their calculation. Second, and more importantly, the present
framework employs fully self-consistent Skyrme-Hartree-Fock plus
RPA calculations with the SLy5 interaction, incorporating all
natural-parity states with multipolarities
$L = 0^+, 1^-, 2^+, 3^-, 4^+, 5^-$ and excitation energies up
to 50~MeV. In contrast,
Ref.~\cite{mau1976optical} used a phenomenological zero-range
force with Soper exchange mixture and a more restricted set of
RPA states, without self-consistency between the mean field and
the residual interaction. The residual difference in the surface non-local range
may therefore reflect the limited model space and the particular choice of
interaction parameters of Ref.~\cite{mau1976optical} rather than a genuine
physical effect; the overall magnitude of the surface non-locality obtained
in the two calculations is, however, in reasonable agreement.
Despite these quantitative differences, both calculations share
the fundamental qualitative result that the imaginary part of
the non-local potential exhibits a Gaussian dependence on the
non-locality coordinate $s$, confirming that this functional form
is independent of the specific form factors of
the underlying zero-range interaction.

The neutron-rich isotope $^{48}$Ca reveals a more complex behavior.
At very low energies ($E < 7$~MeV), the volume parameter
$\beta_{\mathrm{v}}$ at $R = 3.0$~fm exceeds the surface
parameter $\beta_{\mathrm{s}}$ at $R = 4.9$~fm, with
$\beta_{\mathrm{v}}$ reaching approximately $1.60$%
~fm.
With increasing energy, $\beta_{\mathrm{v}}$ decreases rapidly
to $\approx 1.13$--$1.25$~fm, while $\beta_{\mathrm{s}}$ stabilizes
around $1.27$--$1.32$~fm, leading to a crossover at
$E \approx 7$--$8$~MeV above which $\beta_{\mathrm{s}}$
consistently exceeds $\beta_{\mathrm{v}}$. Furthermore,
$\beta_{\mathrm{s}}$ exhibits a non-monotonic structure with a
local maximum around $E \approx 18$--$20$~MeV reaching
$\approx 1.31$--$1.36$~fm before decreasing at higher energies.
This non-trivial energy dependence may be attributed to the neutron
excess in $^{48}$Ca, which modifies the spectrum of collective
excitations and introduces additional isovector modes that couple
differently to the scattering nucleon in the interior and at the
surface. The crossing of $\beta_{\mathrm{v}}$ and
$\beta_{\mathrm{s}}$ as a function of energy is a distinctive
feature of this neutron-rich system. As shown in Fig.~\ref{h3},
the difference $\Delta\beta = \beta_{\mathrm{s}} -
\beta_{\mathrm{v}}$ remains modest, ranging from $\approx 0.07$~fm at
$E = 11$~MeV to $\approx 0.15$~fm at $E = 20$~MeV and $\approx 0.09$~fm at
$E = 30$~MeV.

For the heavy nucleus $^{208}$Pb, the most striking features
among all investigated targets are observed. As shown in
Fig.~\ref{h5}, the volume parameter $\beta_{\mathrm{v}}$ at
$R = 5.7$~fm exhibits a strong energy dependence, starting at
approximately $2.03$~fm at $E \approx 5$~MeV, displaying
pronounced fluctuations with peaks reaching $\approx 1.98$~fm near
$E \approx 10$--$15$~MeV, and then decreasing steeply to
$\approx 1.37$~fm at $E = 30$~MeV. In contrast, the surface parameter
$\beta_{\mathrm{s}}$ at $R = 7.6$~fm shows a non-monotonic
behavior: it starts at $\approx 1.65$~fm at $E \approx 5$~MeV,
exhibits a dip around $E \approx 8$--$9$~MeV, recovers to
$\approx 1.54$--$1.58$~fm at intermediate energies
($E \approx 15$--$20$~MeV), and then settles around
$\approx 1.46$--$1.47$~fm at $E = 30$~MeV. Despite these
fluctuations, $\beta_{\mathrm{s}}$ remains considerably more
stable than $\beta_{\mathrm{v}}$ over the full energy range.
Remarkably, at low energies $\beta_{\mathrm{v}}$ significantly
exceeds $\beta_{\mathrm{s}}$. The crossing between
$\beta_{\mathrm{v}}$ and $\beta_{\mathrm{s}}$ occurs around
$E \approx 22$--$22.5$~MeV, above which the conventional hierarchy
$\beta_{\mathrm{s}} > \beta_{\mathrm{v}}$ is restored.

We next compare in detail with the seminal work of V. Bernard and N. Van~Giai~\cite{bernard1979microscopic} on nucleon-$^{208}$Pb scattering. In
Ref.~\cite{bernard1979microscopic}, the imaginary part of the
optical potential was calculated using the Skyrme SIII interaction
within a framework where the same effective force was used to
generate the Hartree-Fock field, the RPA excited states, and the
particle-vibration coupling. The non-locality of $W(R,s)$ was
analyzed at the peak radius $R = 7.2$~fm, where a Gaussian
shape was found with $\beta_{\mathrm{s}}$ varying slightly from
$0.98$~fm at $E = 5$~MeV to $0.92$~fm at $E = 30$~MeV.
Crucially, in Ref.~\cite{bernard1979microscopic} the absorptive
potential was found to vanish in the nuclear interior, making it
impossible to extract a volume non-locality parameter
$\beta_{\mathrm{v}}$. This absence of interior absorption was
attributed to two reinforcing effects: the strong density
dependence of the SIII coupling strength $\bar{v}(r)$, which
becomes very small inside the nucleus, and the surface-peaked
nature of the transition densities $\rho_N(r)$ for the dominant
low-lying isoscalar collective states. Furthermore, the
calculation of Ref.~\cite{bernard1979microscopic} excluded
spin-orbit, velocity-dependent ($t_{1},t_{2}$), and spin-dependent terms in the
residual interaction, and was restricted to a limited set of
vibrational modes (isoscalar states with $L = 0^+, 2^+, 3^-,
4^+, 5^-$ and isovector states with $L = 0^+, 1^-, 2^+$), which
may have further limited the completeness of the coupling scheme
in the nuclear interior.

In contrast, the present calculation based on the SLy5 Skyrme
interaction with fully self-consistent Hartree-Fock plus RPA,
including all natural-parity states with $L = 0^+, 1^-, 2^+,
3^-, 4^+, 5^-$ and excitation energies up to 50~MeV,
reveals significant absorptive strength not only at the surface
but also in the nuclear interior for all incident energies
considered. This is evident from the left panels of Fig.~\ref{h4},
where $W(R, s=0)$ exhibits clear volume contributions centered
around $R \approx 5$--$6$~fm, particularly at higher energies.
The presence of interior absorption now enables, for the first
time in a Skyrme-based PVC framework applied to $^{208}$Pb, the
extraction of the volume non-locality parameter
$\beta_{\mathrm{v}}$.

Regarding the surface non-locality, our extracted values
$\beta_{\mathrm{s}} = 1.63$~fm at $E = 7$~MeV, $1.59$~fm at
$E = 14.5$~MeV, and $1.46$~fm at $E = 30$~MeV
(Table~\ref{table}), evaluated at $R_{\mathrm{s}} = 7.6$~fm, are
systematically larger than the values of $0.98$--$0.92$~fm
reported in Ref.~\cite{bernard1979microscopic} at $R = 7.2$~fm.

Nevertheless, both calculations exhibit the same qualitative
trend of a decreasing $\beta_{\mathrm{s}}$ with increasing
incident energy. The quantitative difference can be traced to
several factors: the use of SLy5 versus SIII, the substantially
enlarged model space of RPA states in the present work, and
the slightly different evaluation radii ($R = 7.6$~fm versus
$7.2$~fm).

\begin{figure}
	\centering
	\includegraphics[width=1\linewidth]{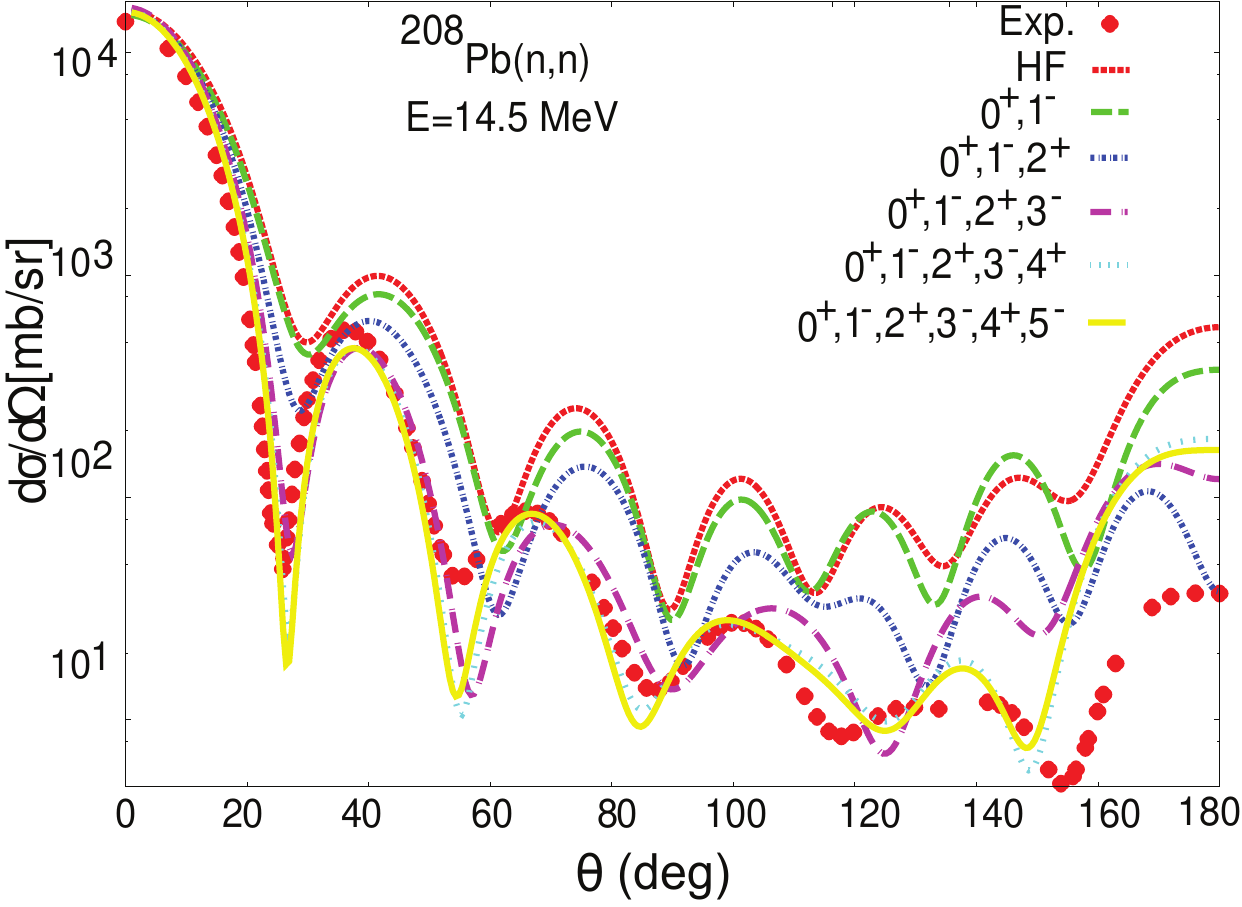}%
	\caption{Contributions of the multipole modes to the angular distribution of neutron elastic scattering by $^{208}$Pb at incident energy $E = 14.5$~MeV. The angular distributions are computed by solving the non-local Schr\"odinger equation exactly with the DWBA98 code~\cite{raynal1998computer}, without introducing any local-equivalent (Perey-type) approximation. The solid line denotes the Hartree--Fock (HF) result. The short-dashed, dotted, long-dashed, dotted/short-dashed, and dotted/long-dashed lines correspond to calculations with maximum multipolarity $L_{\max} = 1, 2, 3, 4$, and $5$, respectively, and the points are the experimental data. Adapted from Ref.~\cite{hao2015low}.}
	\label{fig:obs}
\end{figure}

The most significant advance of the present work for
$^{208}$Pb is the determination of $\beta_{\mathrm{v}}$ at
$R = 5.7$~fm, which was inaccessible in
Ref.~\cite{bernard1979microscopic}. As shown in
Table~\ref{table} and Fig.~\ref{h5}, $\beta_{\mathrm{v}}$ exhibits
a pronounced energy dependence, decreasing from $1.71$~fm at
$E = 7$~MeV to $1.37$~fm at $E = 30$~MeV, a variation of
approximately $20\%$%
. The enhanced volume non-locality at low energies is consistent with a coherent coupling to the low-lying collective excitations of this doubly magic nucleus, an interpretation supported by the multipole decomposition of Fig.~\ref{fig:obs}, which shows that the absorption is built mainly from the most collective $3^{-}$, $2^{+}$, and $4^{+}$ modes. The interior absorptive strength that makes $\beta_{\mathrm{v}}$ accessible (absent in the more restricted SIII calculation of Ref.~\cite{bernard1979microscopic}) originates from the fully self-consistent SLy5 residual interaction and the complete natural-parity RPA model space employed here~\cite{colo2010effect,colo2013self,hao2015low,nhan2018microscopic,hoang2020effects}. Within this interpretation, the interior coupling is strongest at low incident energy, where the superposition of the dominant collective modes produces a broad non-local range extending deep into the nuclear interior and hence a large $\beta_{\mathrm{v}}$; as the incident energy increases and more particle-hole configurations become accessible, this interior coupling is progressively diluted, the interior non-locality is suppressed while the surface non-locality remains comparatively stable, and the two curves eventually cross.
This behavior could not have been anticipated from the surface-only analysis of Ref.~\cite{bernard1979microscopic}, and shows why a description of the absorptive potential over the entire nuclear volume is needed.

Several systematic trends emerge from this comprehensive analysis.
First, at higher energies ($E\gtrsim 20$--$25$~MeV) we find $\beta_{\mathrm{s}} \gtrsim \beta_{\mathrm{v}}$ for all targets. We interpret this ordering as reflecting the surface-peaked character of the transition densities of the dominant collective modes~\cite{bernard1979microscopic}, which localizes the particle-vibration coupling in the surface region where the density gradient is largest. We emphasize, however, that this ordering is \emph{not} universal: for the heavier and neutron-rich systems ($^{48}$Ca and $^{208}$Pb) it is inverted at low energy, where the coherent coupling to interior collective strength makes $\beta_{\mathrm{v}} > \beta_{\mathrm{s}}$. Rather than a fixed hierarchy, the present results thus reveal an energy-dependent competition between surface- and volume-localized non-locality.
Second, the magnitude of $\Delta\beta$ at intermediate energies
varies systematically across the nuclear chart: it is largest for
$^{16}$O ($\approx 0.24$--$0.72$~fm), decreases for $^{40}$Ca
($\approx 0.02$--$0.24$~fm)
and $^{48}$Ca ($\approx 0.07$--$0.15$~fm),
and shows a complex energy-dependent pattern for $^{208}$Pb
(from $\approx -0.38$~fm at $E = 5.5$~MeV to $\approx +0.16$~fm at high
energies)%
, reflecting the interplay between shell structure,
neutron excess, and nuclear density profile. Third, both
$\beta_{\mathrm{v}}$ and $\beta_{\mathrm{s}}$ generally decrease
with increasing incident energy, which can be understood as a
consequence of the diminishing relative importance of low-lying
collective excitations at higher energies: as more particle-hole
configurations become energetically accessible, the coherent
coupling to collective vibrations, which is the primary source
of long-range non-locality, is progressively diluted. This effect
is most dramatic for $\beta_{\mathrm{v}}$ in $^{208}$Pb, where
the variation reaches $\approx 32.5 \%$ from the peak value to the
highest energy considered. Finally, for all targets and energies
considered, the obtained $\beta$ values lie in the range
$1.01$--$2.03$~fm.

\section{Summary and conclusions}
We have studied the non-locality of the imaginary part of a microscopic optical potential that links the structure of the target to the reaction dynamics at low energies, for neutron elastic scattering on $^{16}$O, $^{40}$Ca, $^{48}$Ca, and $^{208}$Pb. Instead of a constant value fitted to experimental data, the extracted non-locality parameter is a radius- and energy-dependent function $\beta(R,E)$, whose values in both the volume and surface regions lie between $1.01$ and $2.03$~fm. Although the present approach is constrained by the use of zero-range Skyrme interactions, the results obtained provide meaningful insights about the non-locality properties that may be useful for the development of next-generation optical potentials applicable across the nuclear chart, from light to heavy systems.

\begin{acknowledgments}
		This research was funded by Hue University under grant no. DHH2025-04-231. The authors (T. V. Nhan Hao and N. Hoang Phuc) also acknowledge the partial support of Hue University under the Core Research Program, Grant No. NCTB.DHH.2025.17. The present research has been also supported by the National Foundation for Science and
		Technology Development of Vietnam (NAFOSTED Project No. 103.04-2025.06).
\end{acknowledgments}

	\bibliography{bibliography}
\end{document}